\documentclass[times]{aastex62}

\graphicspath{{./}{figures/}}

\received{January 1, 2018}
\revised{January 7, 2018}
\accepted{\today}
\submitjournal{AJ}

\shorttitle{Dust Production of Comet 21P During its 2018 Apparition}
\shortauthors{Ehlert, Moticska, and Egal}

\newcommand{\afrho}{$A f \rho$}
\newcommand{\athetafrho}{$A (\theta) f \rho$}
\newcommand{\azerofrho}{$A (0) f \rho$}

\begin{document}

\title{Dust Production of Comet 21P/Giacobini-Zinner Throughout its 2018 Apparition}

\correspondingauthor{S.~Ehlert}
\email{steven.r.ehlert@nasa.gov}

\author[0000-0003-4420-2838]{Steven Ehlert}
\affil{Qualis Corporation, Jacobs Space Exploration Group, NASA Meteoroid Environment Office \\ Marshall Space Flight Center, Huntsville, AL 35812, USA}

\author{Natalie Moticska}
\affil{Physical Sciences Department, Embry-Riddle Aeronautical University \\ Daytona Beach, FL, USA}

\author{Auriane Egal}
\affil{Department of Physics and Astronomy, The University of Western Ontario \\ London, Ontario N6A 3K7, Canada}
\affil{Centre for Planetary Science and Exploration, The University of Western Ontario \\ London, Ontario N6A 5B8, Canada}
\affil{IMCCE, Observatoire de Paris, PSL Research University, CNRS Sorbonne
Universit\'{e}s \\ UPMC Univ. Paris 06, Univ. Lille, France}

\begin{abstract}
We present the results of a long term telescopic observation campaign of comet 21P/Giacobini-Zinner, parent body of the Draconid meteor shower, spanning $\sim 240$ days during its 2018 apparition. Determinations of comet 21P's dust production rate through the \afrho \ parameter derived from these images show that the comet had a highly asymmetric dust production rate that peaked $\sim 10-30$ days before perihelion, when the comet was at a heliocentric distance of $\sim 1.02 - 1.08 \thinspace \mathrm{AU}$. The single highest \afrho \ measurement occurred on 2018-Aug-14 (27 days before perihelion), and had a measured value of \afrho \ = $1594 \thinspace \mathrm{cm}$. The comet's \afrho \ profile is well described by a double-exponential model that rises rapidly during ingress and declines even more rapidly during its egress. These results are fully consistent with observations of comet 21P's dust and gas production rates during past apparitions, and suggest that the double-exponential model we have derived provides a reasonable and stable approximation for the comet's activity over the past thirty to forty years.

\end{abstract}

\keywords{comets: individual Comet 21P/Giacobini-Zinner -- meteoroids -- techniques: photometric }

\section{Introduction}

Of all of the major meteor showers, the October Draconids (009 DRA) remains one of the most difficult to model and forecast. The typical activity of the Draconids is usually comparable to the background sporadic rate for visual observers (with Zenithal Hourly Rates of $\sim 1-2$ per hour, \citep{Kronk2014}), but strong outbursts have been observed in individual years. While some of these outbursts were anticipated, others were entirely unexpected by observers and were not associated with favorable configurations between Earth and the Draconid parent body. Further complicating matters, some of these outbursts were associated only with the lowest mass meteoroids. These outbursts were not observed visually\footnote{By ``visual'' observations, we mean either individual eyewitnesses or optical instruments, which are generally sensitive to meteoroids more massive than $\sim 10^{-5} \thinspace \mathrm{kg}$.}, but were only apparent in radio or radar instrument surveys\footnote{Radar instruments are typically sensitive to masses as low as $\sim 10^{-7} \thinspace \mathrm{kg}$.}. Even for the Draconid outbursts that were anticipated in advance, models have frequently failed to predict the time of peak shower activity, the shower duration, and/or the peak shower flux \citep{Kronk2014}.

Many of the challenges associated with modeling the Draconid meteor shower are associated with the peculiar behavior of the shower's parent body. The parent body of the Draconids, comet 21P/Giacobini-Zinner (hereafter comet 21P) currently has an orbit with a period of 6.54 years, a perihelion distance of 1.0128 AU, and a semi-major axis of 3.50 AU \citep{JPLHorizons}. Its Tisserand parameter with respect to Jupiter is 2.466, classifying it as a Jupiter Family Comet (JFC). Comet 21P has therefore been subject to significant perturbations by Jupiter. Its orbit has been significantly altered over the past century. The most dramatic of these perturbations occurred during its 1959-1965 orbit, over which large changes to its non-gravitational forces were observed \citep{Yeomans1971}.  

Photometric and spectroscopic measurements of the dust and gas production of comet 21P during previous apparitions have shown that this comet has an atypical composition \cite[]{AHearn1995,Lara2003}. In particular, comet 21P is considered the prototypical "carbon-depleted comet", as its $\mathrm{C}_{2}$ abundances have been measured to be significantly lower than the majority of other comets surveyed in \cite{AHearn1995}. Many observers have also provided evidence that gas and dust production for comet 21P reaches its maximum value approximately 1 month prior to its arrival at perihelion. For example, the observations of \cite{Schleicher1987} suggested that gas and dust production dropped by a factor of $\sim 2 $ between pre-perihelion and post-perihelion, a result that was supported by both UV spectroscopy \citep{McFadden1987} and broad-band photometry \citep{Lara2003}. For earlier apparitions, observations suggested that dust production may have peaked after perihelion \citep{Sekanina1985}.

The analysis of \cite{Pittichova2008} showed that during its 2005 apparition, the dust production rate at heliocentric distances of $\sim 1.7-3.0 \thinspace \mathrm{AU}$ (post-perihelion) decreased rapidly during egress, with a logarithmic slope of $\sim -2.0$. When observing the comet at similar heliocentric distances pre-perihelion during its 2011 apparition \citep{Blaauw2014} showed that comet 21P's dust production rate increased even more rapidly during ingress, with a logarithmic slope of $\sim -4.5$. The majority of these results suggest that comet 21P's dust production rises rapidly during ingress, peaks approximately one month before perihelion, and decreases rapidly thereafter. However, discrepancies in coverage, instrumental setups, and measurement techniques between different authors limit this interpretation.

In order to better model the activity of comet 21P and subsequently improve the forecasts for the Draconid meteor shower, NASA's Meteoroid Environment Office (MEO) has undertaken a long-term observation campaign of comet 21P. Unlike observations taken during previous apparitions, we monitored the comet every few nights for nearly three months both before and after perihelion. We utilized telescopes located at multiple locations across the Earth, covering both the Northern and Southern Hemispheres, in order to minimize gaps in between observations. In this work, we present the results of this observation campaign. This is a companion paper to Egal et al. 2019 (submitted, hereafter Egal19) that utilized these observational data in conjunction with improved meteoroid stream models in order to forecast future Draconid meteor showers. Egal19 focuses almost exclusively on modeling the ejection of dust grains from the comet, the dynamics of the meteoroid stream, and comparing the modelled activity of the Draconid shower at Earth each year to historical observations. This paper instead will describe the observational campaign and measurements of comet 21P's dust production.  

This paper is structured as follows: Section \ref{ObsConsiderations} discusses the particulars of the observations including scheduling, telescope hardware, and filters. Section \ref{Measurements} describes the methods by which dust production rates of comet 21P were determined, and Section \ref{Results} presents the measurements themselves. We discuss the implications of these results in the context of meteor shower forecasting and observations of previous apparitions in Section \ref{Discussion}. Unless otherwise noted, all dates and times correspond to UTC.

\section{Observational Considerations}\label{ObsConsiderations}
\begin{figure*}
    \centering
    \includegraphics[width=0.95\textwidth]{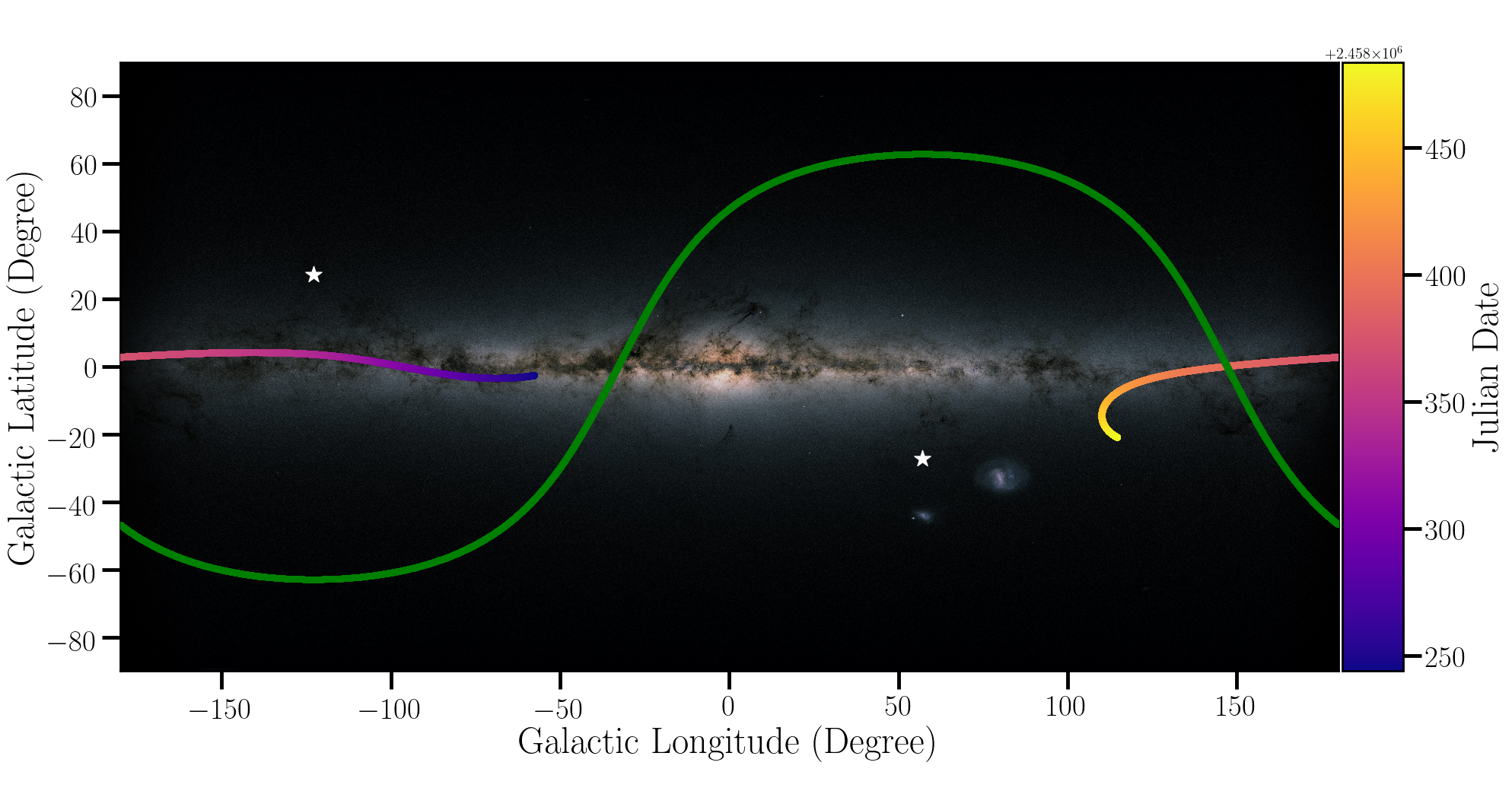}
    \caption{The ephemeris of comet 21P during this observation campaign overlaid on an all-sky image of the Milky Way. The green curve denotes the Celestial Equator ($0^{\circ}$ declination) in Galactic coordinates, with the Northern Hemisphere above the green curve and the Southern Hemisphere below. The North and South celestial poles are denoted by white stars. The color of the ephemeris curve corresponds the Julian Date of the observations, which span from 2018-05-06 through 2018-12-30. The comet's state vector was generally centered on the Galactic Plane throughout the entirety of this observation campaign. The Galactic image is attributed to ESA/Gaia/DPAC.   }
 \label{fig:CometTrajectory}
\end{figure*}

Observations of comet 21P commenced on 2018-05-06, and the final images were taken on 2018-12-30. Figure \ref{fig:CometTrajectory} illustrates the ephemeris of the comet during this time period. From this figure, two crucial aspects of the observation planning are noted. First, the comet was positioned in both the Northern and Southern hemispheres of the sky during this campaign.  Telescopes in both hemispheres were necessarily utilized, and had to accommodate for the comet's non-sidereal motion. Second, the comet was usually located coincident with the Galactic Plane during these observations, meaning that it was usually positioned among crowded star fields. 

\subsection{Telescopes Utilized}

Images of comet 21P were taken using telescopes publicly available through the iTelescope network\footnote{\url{https://www.itelescope.net}}. In particular, we imaged the comet using the following telescopes: T11 in Mayhill, New Mexico; T7 in Nerpio, Spain; and T30/T31 at Siding Spring Observatory in Australia. Table \ref{tab:Telescopes} describes the general properties of each of these telescopes.




\begin{table*}[h]
    \centering
    \begin{tabular}{c c  c c c}
    \hline
         Telescope & Location & Diameter (m) &  Field of View (\arcmin) & Pixel Size (\arcsec)  \\
         \hline\hline
        T7 & Nerpio, Spain &  0.41 &  $  42.3 \times 28.2$ & 1.26\\
         T11 & Mayhill, NM, USA &  0.50 & $54.3 \times 36.2$ & 1.62\\
         T30 & Siding Spring, Australia &  0.50 & $41.6 \times 27.8 $ & 1.62\\
         T31 & Siding Spring, Australia  &  0.50 & $55.9 \times 55.9$ & 2.20\\
         \hline
    \end{tabular}
    \caption{A summary of the telescopes utilized during the course of this observation campaign. The columns denote: 1) the name of the telescope; 2) the location of the observatory where the telescope is located; 3) the diameter of the telescope's primary mirror, in m; 4) the field of view of the telescope, in arcminutes; and 5) the effective pixel size of the telescope camera after binning, in arcseconds.   }
    \label{tab:Telescopes}
\end{table*}

\subsection{Filters and Timing}
All images of comet 21P during this observation campaign were taken using the Johnson-Cousins $R_{C}$-band filter (hereafter simply the $R$-band) in place, the standard filter for measuring \afrho. This filter has an effective wavelength of $6407 \mathrm{\AA}$ and an equivalent width of $1580 \mathrm{\AA}$ \citep{Bessel2005}. Binning was always set to two pixels. Exposure times ranged from fifteen seconds to one minute, in order to ensure that the comet never traveled more than one binned pixel during the exposure. Binning by two also increased the signal-to-noise ratio per pixel and more appropriately sampled the typical seeing at these three observatory sites. 

Because the comet was usually located within crowded star fields, the images were scheduled specifically to ensure separation between the comet and nearby stars. We did this by comparing the position of the comet as determined by JPL Horizons at a given time to the URAT-1 \citep{Zacharias2015} and UCAC-4 \citep{Zacharias2012} stellar catalogs, considering all stars with magnitudes brighter than $19$. We only scheduled observations for times when the angular separation between the comet position and the nearest star was at least $20^{\prime \prime}$ for fifteen minutes. 

Ten images were scheduled for each night, and all ten images were first inspected visually to ensure a minimum quality threshold. Images that were obviously unsuitable for further analysis due to tracking or weather issues were immediately rejected. Bias, dark, and flat-field corrections to each image were done in a standard method automatically by iTelescope. Summaries of the observations taken before and after perihelion can be found in Tables \ref{tab:obs_preperi} and \ref{tab:obs_postperi}, respectively. 

We acquired high quality imaging data of the comet on a total of 70 nights - 32 of which were taken before perihelion and 38 afterwards. The average spacing between consecutive observations is $\sim 3-4$ days, although the cadence is not uniform. Due to poor weather at the observing sites, there were several time periods when images were not taken for 10-14 days. 

\begin{deluxetable}{cccccccc}
\tablecaption{Observations of Comet 21P taken prior to perihelion. The columns denote: 1) The start date and time of the first image taken for each observing session; 2) the telescope utilized for these images (see Table \ref{tab:Telescopes}); 3) the exposure time of each image, in seconds; 4) the right ascension of the comet at the time of observation; 5) the declination of the comet at the time of observation; 6) the heliocentric distance of the comet, in AU; 7) the geocentric distance of the comet, in AU; and 8) the Sun - Comet - Earth phase angle at the time of observation, in degrees. \label{tab:obs_preperi}}
\tablehead{
\colhead{Date}  & 
\colhead{Telescope} & 
\colhead{Exposure (s)} &
\colhead{RA} &
\colhead{DEC} & 
\colhead{$\mathrm{r}_{\mathrm{H}}$}  &  
\colhead{$\Delta$}  &
\colhead{Phase Angle ($^{\circ}$)}}

\startdata
2018-05-06 09:26:36 UT & T11 & 300 & 19$^{\mathrm h}$49$^{ \mathrm m}$47\fs28 & +20\degr 50\arcmin 08.9\arcsec & 1.9188 & 1.4965 & 31.37 \\
2018-05-14 09:24:13 UT & T11 & 60 & 20$^{\mathrm h}$01$^{ \mathrm m}$55\fs58 & +24\degr 26\arcmin 29.2\arcsec & 1.8450 & 1.3821 & 32.68 \\
2018-05-17 08:36:47 UT & T11 & 60 & 20$^{\mathrm h}$06$^{ \mathrm m}$27\fs72 & +25\degr 50\arcmin 48.5\arcsec & 1.8176 & 1.3416 & 33.21 \\
2018-05-24 10:17:02 UT & T11 & 60 & 20$^{\mathrm h}$17$^{ \mathrm m}$24\fs46 & +29\degr 19\arcmin 58.6\arcsec & 1.7520 & 1.2493 & 34.57 \\
2018-05-25 10:17:23 UT & T11 & 60 & 20$^{\mathrm h}$18$^{ \mathrm m}$58\fs72 & +29\degr 50\arcmin 27.9\arcsec & 1.7427 & 1.2367 & 34.78 \\
2018-05-28 07:37:01 UT & T11 & 60 & 20$^{\mathrm h}$23$^{ \mathrm m}$33\fs47 & +31\degr 19\arcmin 41.2\arcsec & 1.7159 & 1.2011 & 35.40 \\
2018-06-05 08:22:15 UT & T11 & 60 & 20$^{\mathrm h}$36$^{ \mathrm m}$43\fs20 & +35\degr 36\arcmin 17.2\arcsec & 1.6414 & 1.1069 & 37.32 \\
2018-06-09 08:12:32 UT & T11 & 60 & 20$^{\mathrm h}$43$^{ \mathrm m}$35\fs13 & +37\degr 47\arcmin 57.9\arcsec & 1.6045 & 1.0627 & 38.39 \\
2018-06-11 09:56:31 UT & T11 & 60 & 20$^{\mathrm h}$47$^{ \mathrm m}$15\fs53 & +38\degr 57\arcmin 11.6\arcsec & 1.5854 & 1.0404 & 38.97 \\
2018-06-12 10:07:35 UT & T11 & 60 & 20$^{\mathrm h}$49$^{ \mathrm m}$04\fs60 & +39\degr 31\arcmin 03.1\arcsec & 1.5761 & 1.0297 & 39.26 \\
2018-06-17 08:44:00 UT & T11 & 60 & 20$^{\mathrm h}$58$^{ \mathrm m}$20\fs87 & 42\degr 18\arcmin 39.0\arcsec & 1.5308 & 0.9788 & 40.78 \\
2018-06-20 05:21:49 UT & T11 & 60 & 21$^{\mathrm h}$04$^{ \mathrm m}$02\fs44 & +43\degr 56\arcmin 32.6\arcsec & 1.5048 & 0.9503 & 41.72 \\
2018-06-30 02:15:03 UT & T7 & 60 & 21$^{\mathrm h}$26$^{ \mathrm m}$24\fs35 & +49\degr 36\arcmin 49.4\arcsec & 1.4164 & 0.8572 & 45.32 \\
2018-07-01 09:17:14 UT & T11 & 60 & 21$^{\mathrm h}$29$^{ \mathrm m}$45\fs00 & 50\degr 21\arcmin 23.8\arcsec & 1.4050 & 0.8455 & 45.84 \\
2018-07-09 02:07:53 UT & T7 & 60 & 21$^{\mathrm h}$52$^{ \mathrm m}$33\fs74 & +54\degr 44\arcmin 17.8\arcsec & 1.3386 & 0.7780 & 49.11 \\
2018-07-12 22:22:46 UT & T7 & 60 & 22$^{\mathrm h}$06$^{ \mathrm m}$21\fs95 & +56\degr 52\arcmin 12.0\arcsec & 1.3065 & 0.7454 & 50.87 \\
2018-07-15 22:22:34 UT & T7 & 60 & 22$^{\mathrm h}$18$^{ \mathrm m}$38\fs49 & +58\degr 29\arcmin 17.0\arcsec & 1.2820 & 0.7204 & 52.31 \\
2018-07-18 05:31:39 UT & T11 & 60 & 22$^{\mathrm h}$29$^{ \mathrm m}$07\fs24 & +59\degr 41\arcmin 23.7\arcsec & 1.2637 & 0.7015 & 53.44 \\
2018-07-19 01:38:31 UT & T7 & 60 & 22$^{\mathrm h}$33$^{ \mathrm m}$12\fs32 & +60\degr 07\arcmin 05.7\arcsec & 1.2570 & 0.6947 & 53.86 \\
2018-07-21 09:27:03 UT & T11 & 60 & 22$^{\mathrm h}$45$^{ \mathrm m}$21\fs38 & +61\degr 16\arcmin 22.9\arcsec & 1.2390 & 0.6759 & 55.05 \\
2018-07-22 22:37:43 UT & T7 & 60 & 22$^{\mathrm h}$54$^{ \mathrm m}$10\fs76 & +62\degr 00\arcmin 36.2\arcsec & 1.2271 & 0.6635 & 55.86 \\
2018-07-24 03:05:18 UT & T7 & 60 & 23$^{\mathrm h}$01$^{ \mathrm m}$21\fs82 & +62\degr 33\arcmin 13.0\arcsec & 1.2182 & 0.6540 & 56.48 \\
2018-07-24 22:32:43 UT & T7 & 60 & 23$^{\mathrm h}$06$^{ \mathrm m}$30\fs16 & +62\degr 54\arcmin 49.0\arcsec & 1.2122 & 0.6476 & 56.92 \\
2018-07-26 01:57:41 UT & T7 & 60 & 23$^{\mathrm h}$14$^{ \mathrm m}$04\fs35 & +63\degr 24\arcmin 11.3\arcsec & 1.2038 & 0.6386 & 57.53 \\
2018-08-05 08:21:41 UT & T11 & 60 & 00$^{\mathrm h}$41$^{ \mathrm m}$25\fs95 & +66\degr 22\arcmin 56.5\arcsec & 1.1341 & 0.5602 & 63.30 \\
2018-08-06 10:43:20 UT & T11 & 60 & 00$^{\mathrm h}$52$^{ \mathrm m}$49\fs76 & +66\degr 28\arcmin 59.2\arcsec & 1.1274 & 0.5521 & 63.94 \\
2018-08-08 06:46:37 UT & T11 & 60 & 01$^{\mathrm h}$12$^{ \mathrm m}$34\fs43 & +66\degr 31\arcmin 20.4\arcsec & 1.1164 & 0.5388 & 65.00 \\
2018-08-14 08:26:57 UT & T11 & 30 & 02$^{\mathrm h}$21$^{ \mathrm m}$10\fs06 & +65\degr 17\arcmin 54.3\arcsec & 1.0834 & 0.4969 & 68.51 \\
2018-08-15 09:02:05 UT & T11 & 30 & 02$^{\mathrm h}$32$^{ \mathrm m}$43\fs46 & +64\degr 51\arcmin 44.8\arcsec & 1.0784 & 0.4902 & 69.09 \\
2018-08-27 10:22:12 UT & T11 & 15 & 04$^{\mathrm h}$29$^{ \mathrm m}$01\fs41 & +54\degr 28\arcmin 57.1\arcsec & 1.0322 & 0.4238 & 75.15 \\
2018-08-31 09:27:18 UT & T11 & 15 & 04$^{\mathrm h}$57$^{ \mathrm m}$08\fs17 & +49\degr 09\arcmin 39.9\arcsec & 1.0228 & 0.4088 & 76.58 \\
2018-09-01 08:21:49 UT & T11 & 15 & 05$^{\mathrm h}$03$^{ \mathrm m}$12\fs99 & +47\degr 45\arcmin 55.2\arcsec & 1.0209 & 0.4059 & 76.86 \\
\enddata
\end{deluxetable}

\begin{deluxetable}{cccccccc}
\tablecaption{Observations of Comet 21P taken after perihelion. The columns are identical to those in Table \ref{tab:obs_preperi}. \label{tab:obs_postperi}}
\tablehead{
\colhead{Date}  & 
\colhead{Telescope} & 
\colhead{Exposure (s)} &
\colhead{RA} &
\colhead{DEC} & 
\colhead{$\mathrm{r}_{\mathrm{H}}$}  &  
\colhead{$\Delta$}  &
\colhead{Phase Angle ($^{\circ}$)}}

\startdata
2018-09-10 09:33:04 UT & T11 & 15 & 05$^{\mathrm h}$50$^{ \mathrm m}$01\fs97 & +33\degr 00\arcmin 54.0\arcsec & 1.0128 & 0.3918 & 77.99 \\
2018-09-12 09:57:17 UT & T11 & 15 & 05$^{\mathrm h}$58$^{ \mathrm m}$17\fs73 & +29\degr 31\arcmin 08.1\arcsec & 1.0133 & 0.3923 & 77.83 \\
2018-09-13 09:02:16 UT & T11 & 15 & 06$^{\mathrm h}$02$^{ \mathrm m}$00\fs79 & +27\degr 50\arcmin 49.6\arcsec & 1.0138 & 0.3931 & 77.70 \\
2018-09-14 11:01:23 UT & T11 & 15 & 06$^{\mathrm h}$06$^{ \mathrm m}$02\fs29 & +25\degr 58\arcmin 02.2\arcsec & 1.0146 & 0.3943 & 77.52 \\
2018-09-18 11:06:48 UT & T11 & 15 & 06$^{\mathrm h}$19$^{ \mathrm m}$35\fs76 & +19\degr 06\arcmin 25.0\arcsec & 1.0197 & 0.4021 & 76.49 \\
2018-09-19 10:26:18 UT & T11 & 15 & 06$^{\mathrm h}$22$^{ \mathrm m}$36\fs17 & +17\degr 28\arcmin 43.0\arcsec & 1.0214 & 0.4047 & 76.16 \\
2018-09-21 10:57:54 UT & T11 & 15 & 06$^{\mathrm h}$28$^{ \mathrm m}$32\fs40 & +14\degr 09\arcmin 21.3\arcsec & 1.0255 & 0.4111 & 75.40 \\
2018-09-25 10:56:50 UT & T11 & 15 & 06$^{\mathrm h}$39$^{ \mathrm m}$08\fs32 & +07\degr 53\arcmin 50.1\arcsec & 1.0360 & 0.4269 & 73.60 \\
2018-09-28 09:47:28 UT & T11 & 15 & 06$^{\mathrm h}$46$^{ \mathrm m}$05\fs59 & +03\degr 35\arcmin 06.6\arcsec & 1.0457 & 0.4409 & 72.08 \\
2018-09-29 10:02:27 UT & T11 & 15 & 06$^{\mathrm h}$48$^{ \mathrm m}$19\fs12 & +02\degr 10\arcmin 22.9\arcsec & 1.0494 & 0.4461 & 71.53 \\
2018-09-30 10:02:09 UT & T11 & 15 & 06$^{\mathrm h}$50$^{ \mathrm m}$26\fs80 & +00\degr 48\arcmin 29.8\arcsec & 1.0532 & 0.4514 & 70.98 \\
2018-10-05 10:22:13 UT & T11 & 15 & 07$^{\mathrm h}$00$^{ \mathrm m}$03\fs79 &$-$05\degr 33\arcmin 00.0\arcsec & 1.0748 & 0.4804 & 68.06 \\
2018-10-08 11:47:13 UT & T11 & 15 & 07$^{\mathrm h}$05$^{ \mathrm m}$06\fs74 &$-$09\degr 02\arcmin 37.2\arcsec & 1.0900 & 0.4996 & 66.22 \\
2018-10-10 10:57:19 UT & T11 & 15 & 07$^{\mathrm h}$08$^{ \mathrm m}$02\fs70 &$-$11\degr 08\arcmin 33.2\arcsec & 1.1004 & 0.5124 & 65.03 \\
2018-10-14 17:02:37 UT & T31 & 30 & 07$^{\mathrm h}$13$^{ \mathrm m}$35\fs36 &$-$15\degr 19\arcmin 15.7\arcsec & 1.1249 & 0.5409 & 62.44 \\
2018-10-15 17:37:43 UT & T31 & 30 & 07$^{\mathrm h}$14$^{ \mathrm m}$45\fs87 &$-$16\degr 15\arcmin 26.0\arcsec & 1.1312 & 0.5479 & 61.82 \\
2018-10-21 15:53:22 UT & T31 & 30 & 07$^{\mathrm h}$20$^{ \mathrm m}$22\fs63 &$-$21\degr 11\arcmin 30.7\arcsec & 1.1697 & 0.5889 & 58.29 \\
2018-10-24 16:43:04 UT & T31 & 30 & 07$^{\mathrm h}$22$^{ \mathrm m}$28\fs72 &$-$23\degr 25\arcmin 39.1\arcsec & 1.1909 & 0.6101 & 56.54 \\
2018-10-26 16:41:50 UT & T31 & 30 & 07$^{\mathrm h}$23$^{ \mathrm m}$34\fs78 &$-$24\degr 48\arcmin 11.5\arcsec & 1.2054 & 0.6240 & 55.40 \\
2018-11-04 14:22:59 UT & T30 & 30 & 07$^{\mathrm h}$25$^{ \mathrm m}$44\fs24 &$-$30\degr 05\arcmin 24.2\arcsec & 1.2738 & 0.6853 & 50.59 \\
2018-11-10 17:42:55 UT & T30 & 30 & 07$^{\mathrm h}$24$^{ \mathrm m}$36\fs53 &$-$33\degr 00\arcmin 43.8\arcsec & 1.3240 & 0.7266 & 47.48 \\
2018-11-11 17:37:57 UT & T30 & 30 & 07$^{\mathrm h}$24$^{ \mathrm m}$13\fs71 &$-$33\degr 26\arcmin 03.6\arcsec & 1.3324 & 0.7332 & 47.00 \\
2018-11-25 15:12:21 UT & T31 & 60 & 07$^{\mathrm h}$13$^{ \mathrm m}$41\fs68 &$-$37\degr 51\arcmin 56.4\arcsec & 1.4534 & 0.8243 & 40.67 \\
2018-11-29 13:22:10 UT & T31 & 60 & 07$^{\mathrm h}$09$^{ \mathrm m}$11\fs19 &$-$38\degr 38\arcmin 02.2\arcsec & 1.4886 & 0.8499 & 39.04 \\
2018-12-01 14:32:27 UT & T31 & 60 & 07$^{\mathrm h}$06$^{ \mathrm m}$37\fs14 &$-$38\degr 57\arcmin 03.9\arcsec & 1.5072 & 0.8633 & 38.22 \\
2018-12-04 14:22:00 UT & T31 & 60 & 07$^{\mathrm h}$02$^{ \mathrm m}$38\fs81 &$-$39\degr 18\arcmin 37.8\arcsec & 1.5344 & 0.8831 & 37.05 \\
2018-12-05 14:39:02 UT & T31 & 60 & 07$^{\mathrm h}$01$^{ \mathrm m}$15\fs16 &$-$39\degr 24\arcmin 14.8\arcsec & 1.5437 & 0.8898 & 36.67 \\
2018-12-06 14:10:02 UT & T31 & 60 & 06$^{\mathrm h}$59$^{ \mathrm m}$52\fs90 &$-$39\degr 28\arcmin 52.9\arcsec & 1.5527 & 0.8963 & 36.30 \\
2018-12-07 16:41:41 UT & T31 & 60 & 06$^{\mathrm h}$58$^{ \mathrm m}$18\fs79 &$-$39\degr 33\arcmin 09.8\arcsec & 1.5628 & 0.9037 & 35.89 \\
2018-12-18 15:06:26 UT & T31 & 60 & 06$^{\mathrm h}$42$^{ \mathrm m}$13\fs65 &$-$39\degr 22\arcmin 51.9\arcsec & 1.6637 & 0.9802 & 32.18 \\
2018-12-23 13:42:10 UT & T31 & 60 & 06$^{\mathrm h}$35$^{ \mathrm m}$07\fs87 &$-$38\degr 48\arcmin 47.7\arcsec & 1.7096 & 1.0172 & 30.72 \\
2018-12-24 12:16:50 UT & T31 & 60 & 06$^{\mathrm h}$33$^{ \mathrm m}$49\fs47 &$-$38\degr 40\arcmin 25.3\arcsec & 1.7183 & 1.0244 & 30.46 \\
2018-12-25 16:52:33 UT & T31 & 60 & 06$^{\mathrm h}$32$^{ \mathrm m}$11\fs71 &$-$38\degr 28\arcmin 59.4\arcsec & 1.7293 & 1.0337 & 30.14 \\
2018-12-26 15:02:55 UT & T31 & 60 & 06$^{\mathrm h}$30$^{ \mathrm m}$57\fs18 &$-$38\degr 19\arcmin 30.4\arcsec & 1.7379 & 1.0410 & 29.90 \\
2018-12-27 15:01:35 UT & T31 & 60 & 06$^{\mathrm h}$29$^{ \mathrm m}$37\fs92 &$-$38\degr 08\arcmin 39.3\arcsec & 1.7472 & 1.0490 & 29.64 \\
2018-12-28 16:03:39 UT & T31 & 60 & 06$^{\mathrm h}$28$^{ \mathrm m}$16\fs75 &$-$37\degr 56\arcmin 40.8\arcsec & 1.7569 & 1.0575 & 29.38 \\
2018-12-29 14:03:17 UT & T31 & 60 & 06$^{\mathrm h}$27$^{ \mathrm m}$06\fs85 &$-$37\degr 45\arcmin 37.9\arcsec & 1.7654 & 1.0650 & 29.15 \\
2018-12-30 14:32:47 UT & T31 & 60 & 06$^{\mathrm h}$25$^{ \mathrm m}$50\fs65 &$-$37\degr 32\arcmin 45.7\arcsec & 1.7749 & 1.0734 & 28.91 \\
\enddata
\end{deluxetable}

\section{Image Calibration and Comet Measurements}\label{Measurements}

Astrometric and photometric calibration was performed for every image using two independent pipelines. The first is nearly identical to that presented in \cite{Hosek2013} and \cite{Blaauw2014}, and utilizes the software package \textit{Astrometrica} \citep{Astrometrica} to identify reference stars within the image field of view. The second utilizes \textit{SExtractor} \citep{Bertin1996} to detect and identify reference stars within the field of view. Each pipeline provides the astrometric solution for each image, as well as an image-specific zero-point for photometric calibration. The US Naval Observatory A2 Catalog \citep[][hereafter USNO A2]{Monet1998} was the reference catalog for both pipelines, using their $R$-band magnitudes. 

\subsection{Comet Photometry}

With astrometric and photometric calibration models determined for each image, the magnitude of the comet was then determined. Our dust production measurements were determined using the magnitude of the comet as measured in the $10^{\prime\prime} \times 10^{\prime\prime}$ square aperture centered on the brightest pixel associated with the comet emission. This magnitude was determined by two independent programs: \textit{Fotometria con Astrometrica} \citep[FoCAs,][]{FoCAs} was utilized to ingest the calibration model as derived by \textit{Astrometrica}, whereas an independent \texttt{Python}-based program was utilized to ingest the \textit{SExtractor} derived calibration model. Both programs were confirmed to provide consistent magnitudes for the data presented in \cite{Hosek2013} and \cite{Blaauw2014}. The largest difference between the analyses of \cite{Hosek2013} and \cite{Blaauw2014} and this work is the choice of reference catalog. Both of these older works calibrated their images with respect to the Carlsberg Meridian Catalog, Release 14 \cite[][hereafter CMC-14]{CMC2006}. During this apparition, comet 21P was positioned outside of the CMC-14 survey area for the months of July and August\footnote{Since these papers were completed, Carlsberg Meridian Catalog Release 15 was made public. Neither data release covers regions of the sky with declination of $\delta > 60^{\circ}$, where comet 21P was located for these months.}. The only other reference catalog compatible with \textit{FoCAs} is USNO A2, and we therefore used this catalog throughout the observation campaign. 

The USNO A2 catalog provides $R$-band magnitudes for stars across the entire sky and also provides a useful means for cross-calibrating the measurements by using two separate pipelines. One limitation of this catalog, however, is that reference stars in  different regions of the sky can be subject to rather large calibration discrepancies (of order $\sim 0.3-0.5 \thinspace \mathrm{mag}$), introducing a systematic uncertainty associated with comparing the nightly variations of the comet's magnitude (and subsequently \afrho \ measurements) as it moves across the sky\footnote{This level of plate-to-plate variation was documented in the for the USNO A1 catalog (\url{http://vizier.cfa.harvard.edu/vizier/VizieR/pmm/usno2.htx\#pht}). Although the USNO A2 analysis procedure made efforts to smooth out the variance between plates, the extent to which systematic uncertainties were reduced was not quantified. }.

In addition to the magnitudes used to measure dust production, we also present total coma magnitudes for each night. These magnitudes were determined in circular apertures. The radius of this total coma aperture was determined by visual inspection of the images from each night, and varied from 5-21 pixels. The aperture radius associated with the total coma was determined by visually identifying a sharp edge in the coma's surface brightness in the direction of Sun, and choosing the smallest integer radius that encloses this edge. The corresponding magnitudes were only determined using the \textit{SExtractor}-based pipeline.

\subsection{Determining \afrho \ from Comet Photometry}

\begin{figure}
    \centering
    \includegraphics[width=0.5\columnwidth]{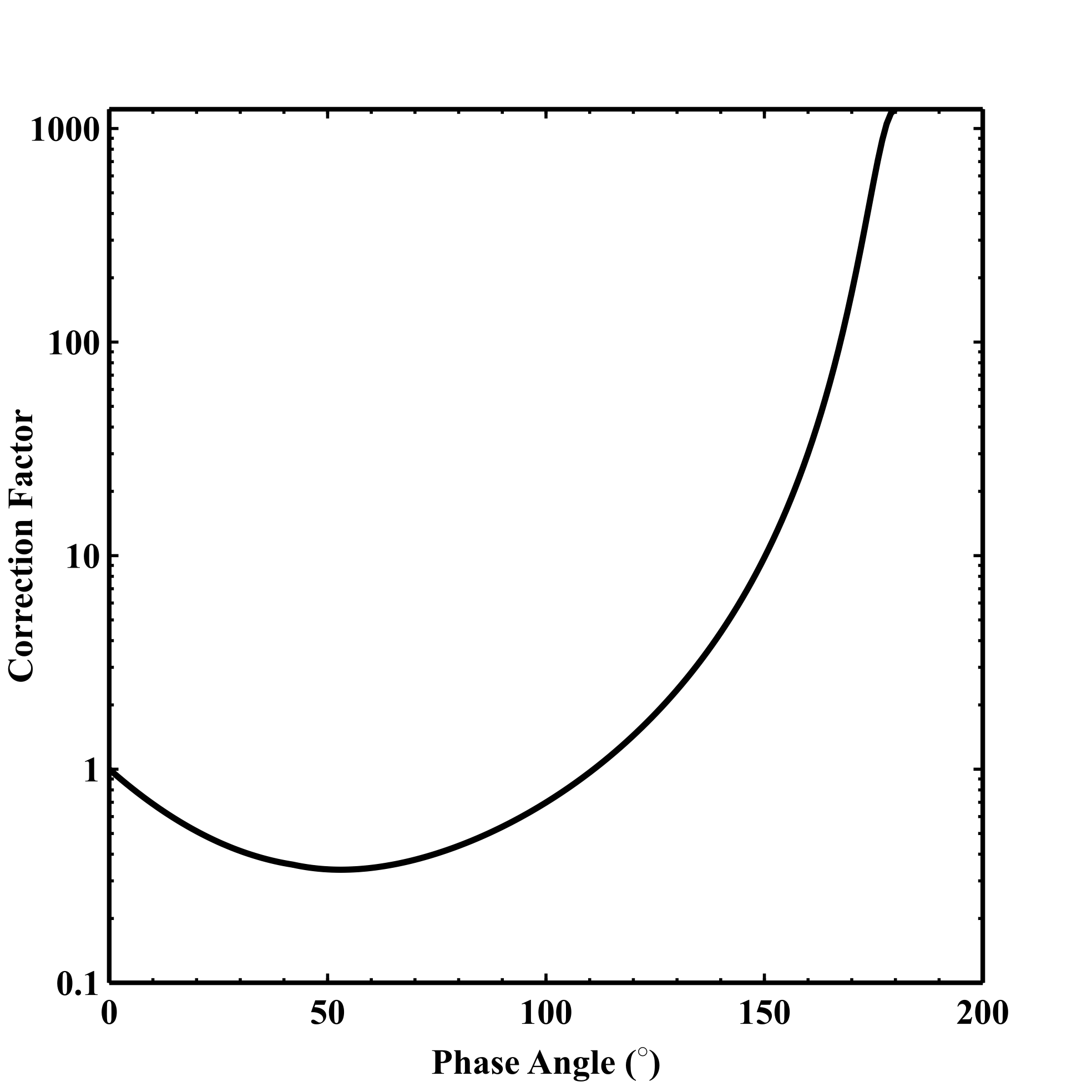}
    \caption{The Schleicher phase function assumed for accounting for variations in observing geometry between different nights. This phase function is normalized to have a value of unity at $0^{\circ}$.  }
    \label{fig:phasefunction}
\end{figure}

With photometric magnitudes of the comet measured for each image, we then convert these magnitudes into dust production rates using the \afrho \ parameter described by \cite{AHearn1984}. As discussed in this paper, \afrho \ is calculated from the comet's measured photometric flux $F$ within a circular aperture of radius $\rho$ (in cm) as

\begin{equation}
    A(\theta) f \rho = \frac{4}{\rho} \times \Delta^{2} \times  \mathrm{r}_{\mathrm{H}}^{2} \times \left(\frac{F}{F_{\bigodot}}\right)
    \label{eq:afrho}
\end{equation}
\noindent where $\mathrm{r}_{\mathrm{H}}$ (in AU) and $\Delta$ (in cm) correspond to the heliocentric and geocentric distances to the comet, respectively. The photometric flux of the Sun is given as $F_{\bigodot}$. For this work, we use the same calculations as \cite{Hosek2013} and \cite{Blaauw2014}. The effective radius $\rho$ of the square aperture is given as 

\begin{equation}
    \rho = \tan\left(\frac{1.12838 r}{206265^{\prime\prime}}\right) \times \Delta
\end{equation}

\noindent where $r = 5^{\prime\prime}$. The comet's photometric flux ratio is calculated from its observed magnitude $R$ as 
\begin{equation}
    \left(\frac{F}{F_{\bigodot}}\right) = 10^\frac{-(R + 27.15)}{2.5}
\end{equation}
\noindent where we assume the apparent $R$-band magnitude of the Sun is -27.15 \citep{Mann2015}. As suggested in Equation \ref{eq:afrho}, the initial value of \afrho \ does not account for how the brightness of the comet varies with phase angle. The initial determination of \athetafrho \ is then corrected for the phase angle using the Schleicher phase function (shown in Figure \ref{fig:phasefunction}), which we will denote as \azerofrho. This phase function is a splice of measurements derived from observations of Comet Halley \citep{Schleicher1998} at low phase angles and a Henyey-Greenstein model at high phase angles \citep{Marcus2007}.  

\section{Results}\label{Results}
\begin{figure*}
    \centering
    \includegraphics[width=0.4\textwidth]{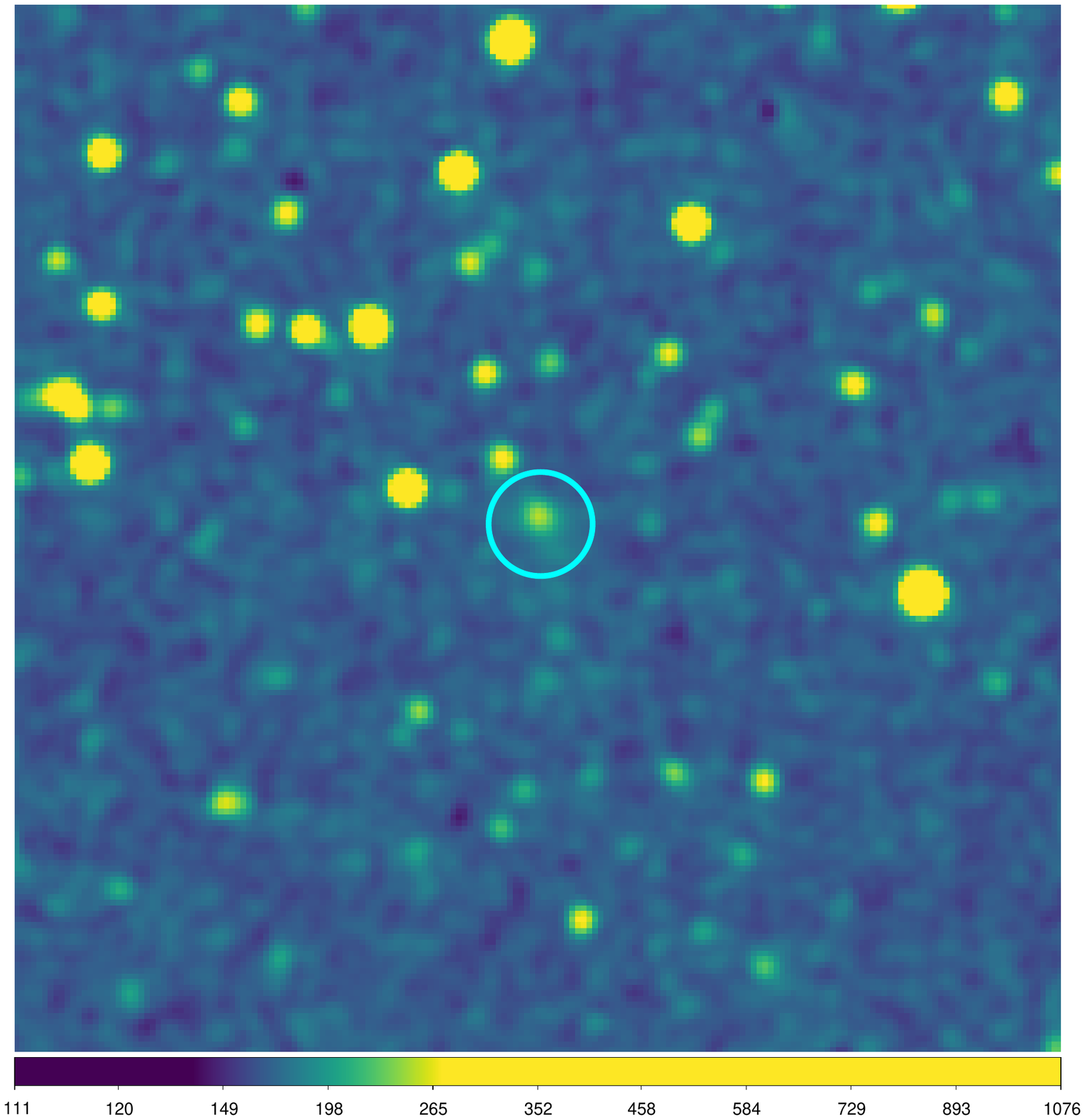}
    \includegraphics[width=0.4\textwidth]{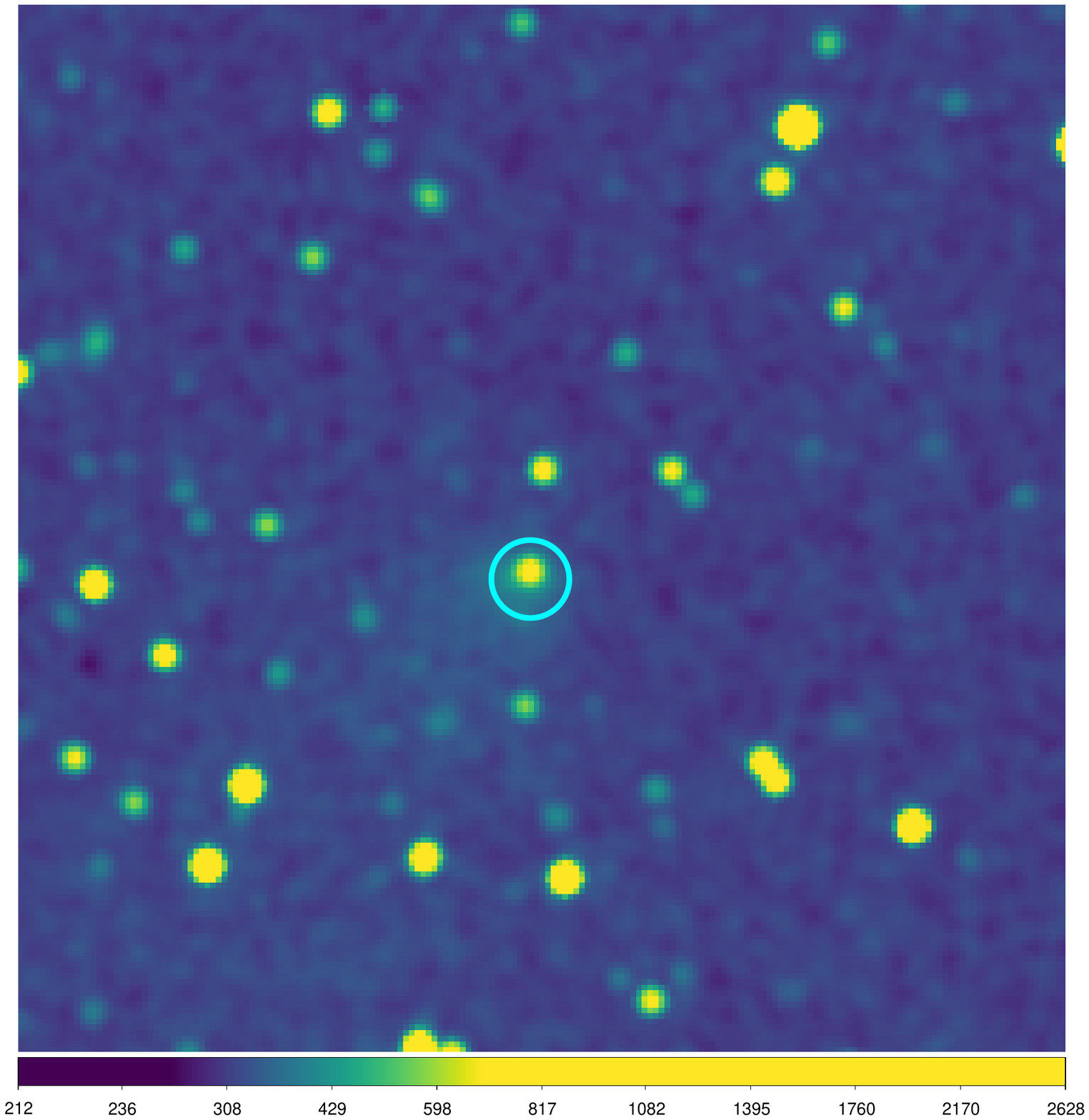}
    \includegraphics[width=0.5\textwidth]{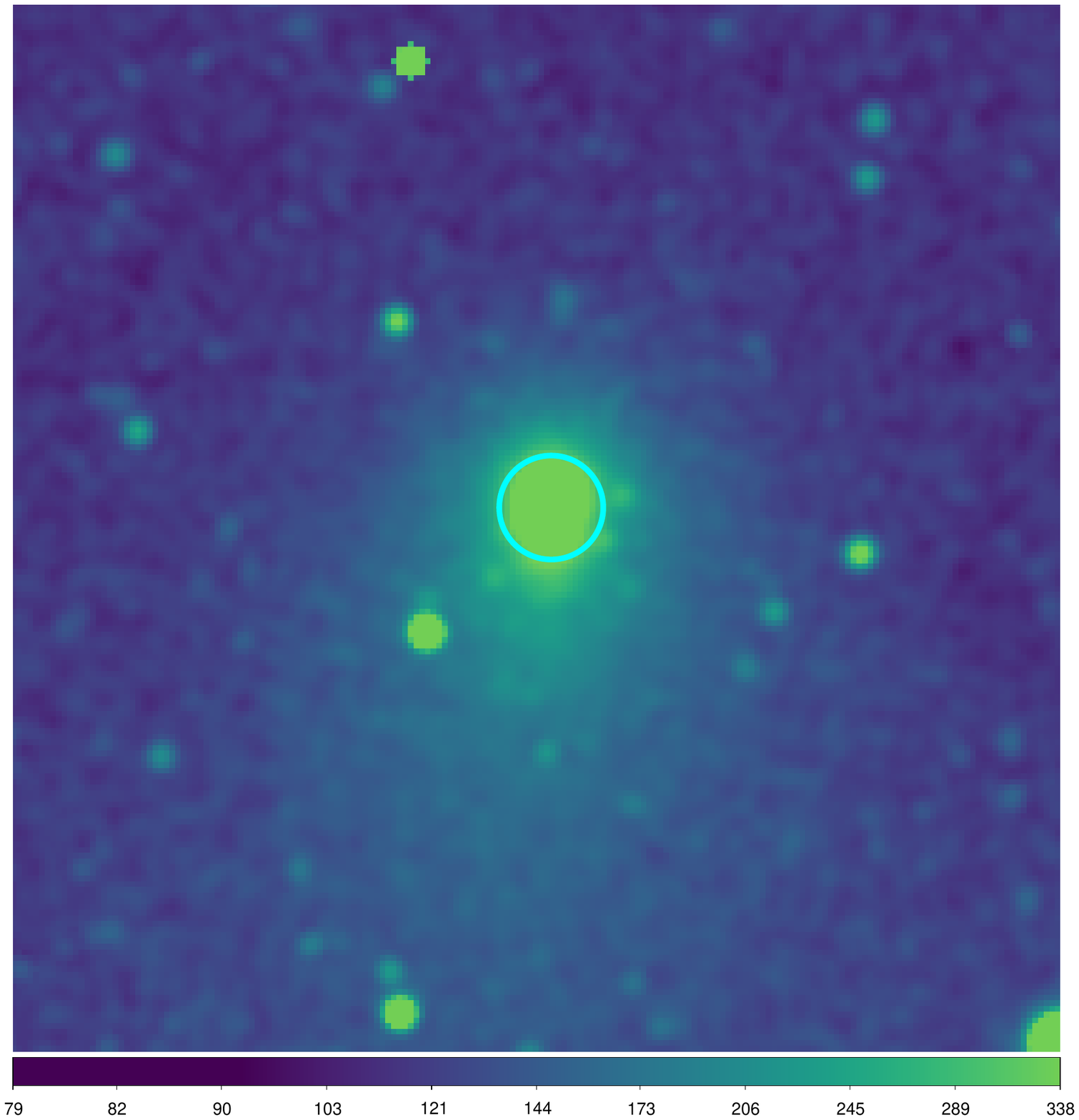}
    \caption{Images of comet 21P as observed at different times during this observation campaign. In all three images the cyan circle denotes a fixed radius of $15\arcsec$. \textit{Top Left}: An image of comet 21P taken on the night of 2018-05-17. The comet is consistent with point-like emission. All images have been smoothed using a Gaussian kernel with a $\sigma =2$ pixels. \textit{Top Right}: An image of comet 21P taken on the night of 2018-12-30, the last night of the observation campaign. The comet emission is once again largely point-like. \textit{Bottom}: An image of comet 21P taken on the night of 2018-09-10, when the comet was at perihelion. In this image, extended emission from the comet's coma is obvious. }
    \label{fig:CometImages}
\end{figure*}

A few example images of comet 21P as observed at the beginning and end of this observing campaign, as well as at perihelion are shown in Figure \ref{fig:CometImages}. The resulting \afrho \ measurements for each night can be found in Tables \ref{tab:afrho_results_pre} and \ref{tab:afrho_results_post}.

\begin{figure*}
    \centering
    \includegraphics[width=0.48\columnwidth]{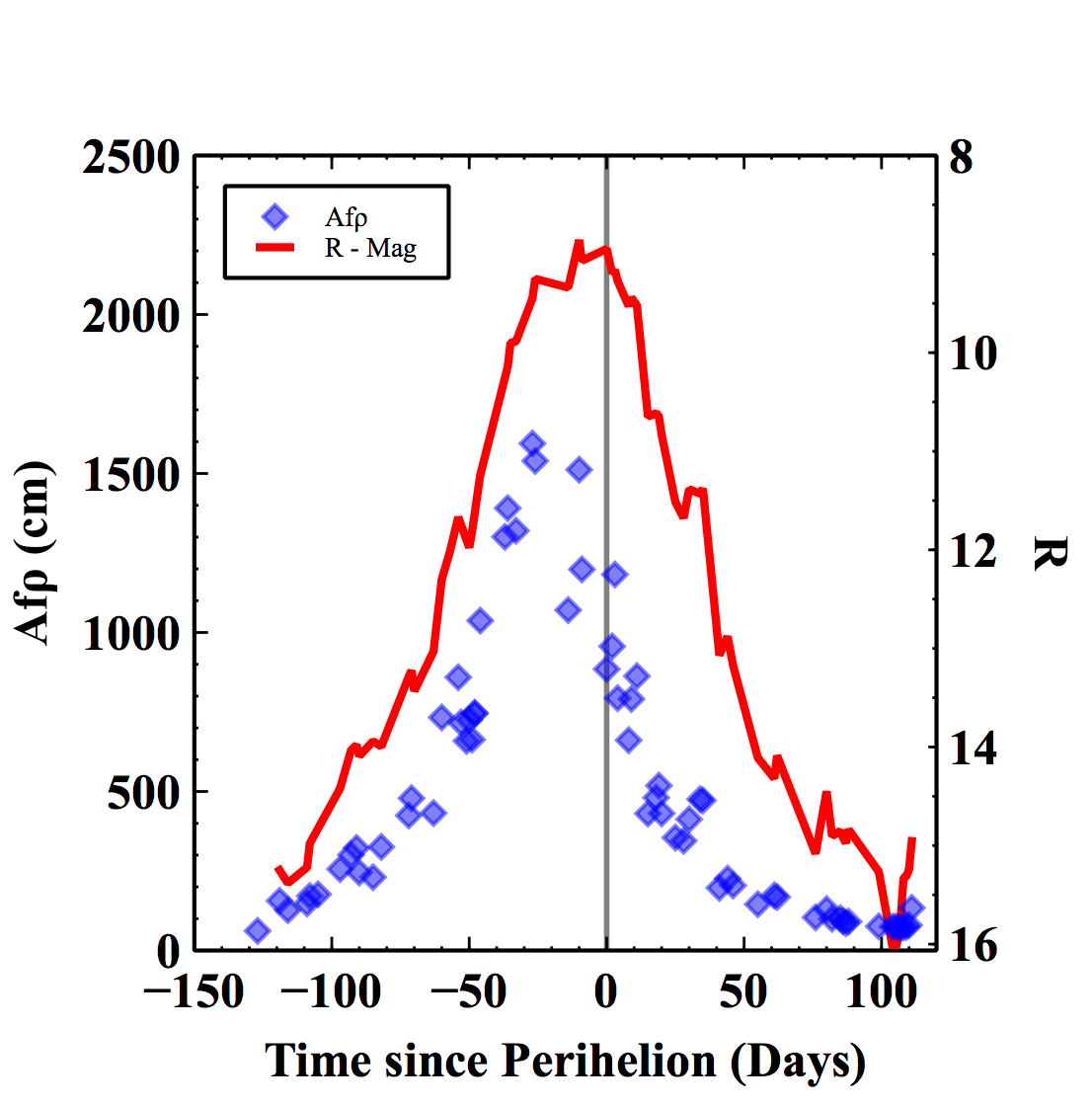}    
     \includegraphics[width=0.48\columnwidth]{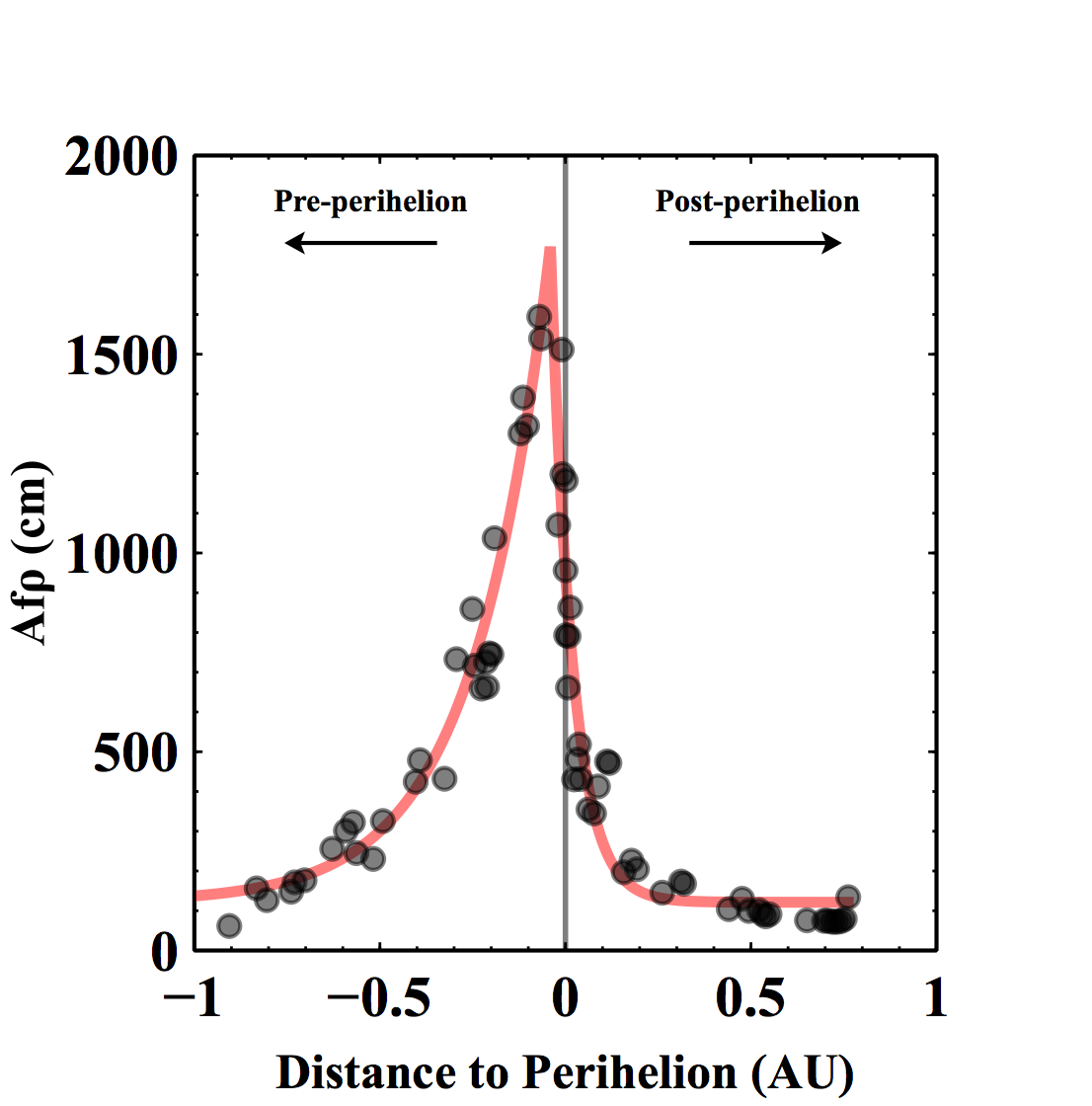}
    \caption{\label{fig:resultplots}The measured values of \afrho \ during the 2018 observation campaign. \textit{Left:} The total apparent magnitude (solid curve) and \azerofrho \ (gray diamonds) of comet 21P as a function of time and heliocentric distance during its 2018 apparition. The date at which the comet reached perihelion, which we used as the reference date, is denoted by the vertical gray line. While the comet reached peak total brightness at perihelion, its dust production rate (as measured by \afrho) peaked approximately $\sim 10-30$ days prior to perihelion.  \textit{Right:}   The measured values of \azerofrho \ of comet 21P as a function of heliocentric distance. The distances have been normalized to define post-perihelion distances as positive and pre-perihelion distances as negative. The solid red curve is the best-fit double-exponential model to these observations from Egal19.}
\end{figure*}

We plot the measured values of \azerofrho \ in two different sets of units. First, in Figure \ref{fig:resultplots}a, we show how the total apparent magnitude and corresponding \azerofrho \ measurement vary with time, using the date of perihelion as a reference date. Second, we show how \azerofrho \ varies with heliocentric distance in Figure \ref{fig:resultplots}b. This plot uses the same re-normalized, signed distances $x$ described in Egal19, where $x$ is defined as 
\begin{equation}
    x=(\mathrm{r}_\mathrm{H}-q)\frac{t - t(q)}{|t-t(q)|}
\end{equation}
\noindent where $q$ is the perihelion distance of comet 21P, $t$ corresponds to the date/time of the observations, and $t(q)$ is the date/time at perihelion. We arbitrarily define pre-perihelion distances in these units to be negative and post-perihelion distances to be positive. 

Both of these figures, in particular Figure \ref{fig:resultplots}b, show that the comet reached its maximum value of \afrho \ before perihelion. Our maximum single value of \afrho \ ($1594 \thinspace \mathrm{cm}$) was measured on the night of 2018-08-14 (27 days before perihelion), although the interpretation of this single measurement is limited by the lack of observational data on neighboring nights and the overall uncertainty in absolute photometric calibration. Similarly high \afrho \ values were measured on 2018-08-15 and 2018-08-31, corresponding to 26 and 10 days before perihelion, respectively. This range of dates corresponds to variations in heliocentric distance between 1.08 AU and 1.02 AU. Unfortunately, few observations of the comet were acquired in the latter half of August and first half of September due to poor weather at the two Northern Hemisphere observation sites. While these measurements provide strong evidence that the maximum value of \afrho \ occurred before perihelion, we cannot state with certainty precisely where within this time window comet 21P reached its maximum value of \afrho. 

These figures also make it clear that comet 21P's \afrho \ values are highly asymmetric around the peak value, with a much steeper descending branch than ascending branch, regardless of whether \afrho \ is considered a function of heliocentric distance or time. In Egal19, the \afrho \ curve in Figure \ref{fig:resultplots}b was fit with a double-exponential model of the form 
\begin{eqnarray}\label{eq:afrhomodel}
Af\rho(x) = \left\{
        \begin{array}{ll}
            K_1+Af\rho(x_{max})*10^{-\gamma_{1} |x-x_{max}|} & \quad x \leq 0 \\
            K_1+Af\rho(x_{max})*10^{-\gamma_{2} |x-x_{max}|} & \quad x \ge 0
        \end{array}
    \right.
\end{eqnarray}

\noindent where $K_1$ is the asymptotic value of \afrho, and $\gamma_1$ and $\gamma_2$ correspond to the logarithmic slopes of the ascending and descending branches, respectively. The best-fit double-exponential model is overlaid on the measurements in Figure \ref{fig:resultplots}b, and has $K_1 = 121$, $x_{max} = 0.04$, $Af\rho(x_{max})=1653$. The best fit logarithmic slopes are $\gamma_1 = 2.10 $ and $\gamma_2 = 7.38$, respectively. This model form better describes the data than other shapes such as a Gaussian, Moffat, or Lorentzian. It also offers more physically sensible model parameters than a broken power-law model fit, as in this latter case the two branches do not agree on a single prediction for the peak value of \afrho.   

\begin{deluxetable}{ccccccc}
\tablecaption{Measurements of \afrho \ for Comet 21P during its 2018 apparition pre-perihelion. The columns denote: 1) The date and time of the observations; 2) the physical size of the $10^{\prime\prime} \times 10^{\prime\prime}$ square aperture, in units of $1000 \thinspace \mathrm{km}$; 3) the extent of the coma used to measure the total magnitude, in units of image pixels (with the corresponding extent in arcseconds given in parentheses); 4) The $R$-band magnitude of the comet as measured within the $10^{\prime\prime} \times 10^{\prime\prime}$ aperture used for \afrho \ measurements; 5) The $R$-band magnitude of the entire coma, as measured in a circular aperture with a radius denoted in column 3; 6) the resultant measurement of \afrho \ before accounting for the phase angle correction, in units of cm; and 7) the final phase-angle corrected values of \afrho, also in units of cm.\label{tab:afrho_results_pre}}
\tablehead{
\colhead{Date}  & 
\colhead{$\rho \thinspace (\times 1000 \thinspace \mathrm{km})$} & 
\colhead{Radius} &
\colhead{$\mathrm{R}_{Af\rho}$} & 
\colhead{$\mathrm{R}_{\mathrm{Coma}}$}  &  
\colhead{\athetafrho \ (cm)}  &
\colhead{\azerofrho \ (cm)}}

\startdata
2018-05-06 & 6.12 & 5 (3.1)   & 17.00 & 17.00 & 26.36 & 61.63 \\
2018-05-14 & 5.65 & 5 (3.1)   & 15.84 & 15.25 & 65.54 & 156.24 \\
2018-05-17 & 5.49 & 5 (3.1)   & 16.01 & 15.38 & 52.71 & 126.64 \\
2018-05-24 & 5.12 & 5 (3.1)   & 15.71 & 15.22 & 60.27 & 147.80 \\
2018-05-25 & 5.06 & 5 (3.1)   & 15.53 & 14.98 & 69.67 & 171.39 \\
2018-05-28 & 4.91 & 5 (3.1)   & 15.44 & 15.28 & 71.11 & 176.56 \\
2018-06-05 & 4.53 & 5 (3.1)   & 14.88 & 14.42 & 100.10 & 255.85 \\
2018-06-09 & 4.20 & 5 (3.1)   & 14.60 & 14.03 & 115.85 & 300.90 \\
2018-06-11 & 4.26 & 5 (3.1)   & 14.52 & 13.98 & 122.99 & 322.27 \\
2018-06-12 & 4.22 & 5 (3.1)   & 14.80 & 14.09 & 93.12 & 245.06 \\
2018-06-17 & 4.01 & 5 (3.1)   & 14.77 & 13.94 & 85.61 & 230.42 \\
2018-06-20 & 3.89 & 7 (4.3)   & 14.34 & 13.99 & 119.35 & 325.67 \\
2018-06-30 & 3.50 & 7 (4.3)   & 13.86 & 13.29 & 147.97 & 424.40 \\
2018-07-01 & 3.46 & 7 (4.3)   & 13.71 & 13.22 & 165.82 & 478.74 \\
2018-07-09 & 3.19 & 7 (5.5)   & 13.66 & 13.03 & 144.08 & 432.02 \\
2018-07-12 & 3.05 & 7 (4.3)   & 13.01 & 12.30 & 240.39 & 732.79 \\
2018-07-15 & 2.95 & 7 (5.5)  & 12.75 & 12.02 & 284.88 & 877.97 \\
2018-07-18 & 2.87 & 9 (7.1)  & 12.72 & 11.67 & 276.85 & 859.17 \\
2018-07-19 & 2.84 & 9 (7.1)  & 12.89 & 11.92 & 230.38 & 716.49 \\
2018-07-21 & 2.77 & 11 (8.7)  & 12.93 & 11.79 & 210.84 & 658.89 \\
2018-07-22 & 2.71 & 11 (6.7)  & 12.79 & 11.98 & 231.61 & 725.33 \\
2018-07-23 & 2.71 & 11 (8.7)  & 12.87 & 11.82 & 211.66 & 663.15 \\
2018-07-24 & 2.68 & 11 (8.7)  & 12.72 & 11.75 & 238.43 & 747.47 \\
2018-07-24 & 2.65 & 11 (6.7)  & 12.71 & 11.75 & 237.46 & 744.69 \\
2018-07-26 & 2.62 & 13 (10.3) & 12.32 & 11.24 & 330.72 & 1037.21 \\
2018-08-05 & 2.29 & 15 (9.2)  & 11.77 & 10.14 & 426.49 & 1300.70 \\
2018-08-06 & 2.26 & 15 (9.2)  & 11.66 & 9.90 & 458.96 & 1391.04 \\
2018-08-08 & 2.20 & 15 (9.2)  & 11.66 & 9.89 & 440.50 & 1319.57 \\
2018-08-14 & 2.03 & 17 (10.4) & 11.25 & 9.45 & 559.07 & 1594.35 \\
2018-08-15 & 2.00 & 17 (10.4) & 11.25 & 9.25 & 544.94 & 1539.27 \\
2018-08-27 & 1.74 & 19 (11.6) & 11.26 & 9.34 & 427.25 & 1071.46 \\
2018-08-31 & 1.68 & 19 (11.6) & 10.79 & 8.86 & 621.60 & 1510.27 \\
2018-09-01 & 1.66 & 19 (11.6) & 11.03 & 9.07 & 496.33 & 1198.40 \\
\enddata
\end{deluxetable}

\begin{deluxetable}{ccccccc}
\tablecaption{Measurements of \afrho \ for Comet 21P during its 2018 apparition post-perihelion. The columns are identical to those of Table \ref{tab:afrho_results_pre}. \label{tab:afrho_results_post}}
\tablehead{
\colhead{Date}  & 
\colhead{$\rho \thinspace (\times 1000 \thinspace \mathrm{km})$} & 
\colhead{Radius} &
\colhead{$\mathrm{R}_{Af\rho}$} & 
\colhead{$\mathrm{R}_{\mathrm{Coma}}$}  &  
\colhead{\athetafrho \ (cm)}  &
\colhead{\azerofrho \ (cm)}}

\startdata
2018-09-10 & 1.60 & 21 (12.8) & 11.27 & 8.95 & 376.07 & 884.72 \\
2018-09-12 & 1.60 & 21 (12.8) & 11.19 & 9.20 & 405.04 & 956.35 \\
2018-09-13 & 1.62 & 21 (12.8) & 10.97 & 9.16 & 499.31 & 1182.39 \\
2018-09-14 & 1.62 & 21 (12.8) & 11.41 & 9.27 & 333.56 & 793.27 \\
2018-09-18 & 1.65 & 19 (11.6) & 11.67 & 9.53 & 271.60 & 661.23 \\
2018-09-19 & 1.66 & 19 (11.6) & 11.49 & 9.44 & 322.30 & 790.42 \\
2018-09-21 & 1.68 & 19 (11.6) & 11.44 & 9.52 & 345.95 & 862.93 \\
2018-09-25 & 1.75 & 15 (9.2)  & 12.30 & 10.65 & 166.04 & 430.47 \\
2018-09-28 & 1.81 & 15 (9.2)  & 12.27 & 10.61 & 179.78 & 480.68 \\
2018-09-29 & 1.83 & 13 (7.9)  & 12.22 & 10.65 & 191.81 & 518.36 \\
2018-09-30 & 1.84 & 13 (7.9)  & 12.45 & 10.83 & 157.89 & 431.19 \\
2018-10-05 & 1.96 & 11 (6.7)  & 12.83 & 11.52 & 123.56 & 354.89 \\
2018-10-08 & 2.05 & 11 (6.7)  & 12.96 & 11.68 & 116.91 & 344.92 \\
2018-10-10 & 2.09 & 11 (6.7)  & 12.83 & 11.38 & 137.76 & 412.58 \\
2018-10-14 & 2.21 & 9 (4.1)   & 12.81 & 11.43 & 154.78 & 475.56 \\
2018-10-15 & 2.24 & 9 (4.1)   & 12.85 & 11.38 & 152.81 & 471.69 \\
2018-10-21 & 2.41 & 7 (3.2)   & 13.97 & 13.07 & 62.79 & 196.77 \\
2018-10-24 & 2.50 & 7 (3.2)   & 13.90 & 12.88 & 72.05 & 225.87 \\
2018-10-26 & 2.54 & 5 (2.3)   & 14.06 & 13.17 & 64.90 & 203.03 \\
2018-11-04 & 2.83 & 5 (3.0)   & 14.62 & 14.11 & 48.15 & 146.42 \\
2018-11-10 & 2.98 & 5 (3.0)   & 14.54 & 14.32 & 58.81 & 173.23 \\
2018-11-11 & 3.01 & 5 (3.0)   & 14.59 & 14.09 & 57.35 & 167.95 \\
2018-11-25 & 3.37 & 5 (3.0)   & 15.33 & 15.08 & 38.69 & 103.97 \\
2018-11-29 & 3.47 & 5 (2.3)   & 15.15 & 14.45 & 49.57 & 130.01 \\
2018-12-01 & 3.53 & 5 (2.3)   & 15.47 & 14.90 & 38.41 & 99.52 \\
2018-12-04 & 3.62 & 5 (2.3)   & 15.49 & 14.85 & 39.91 & 101.59 \\
2018-12-05 & 3.64 & 5 (2.3)   & 15.62 & 14.92 & 36.17 & 91.55 \\
2018-12-06 & 3.67 & 5 (2.3)   & 15.70 & 14.97 & 34.17 & 85.99 \\
2018-12-07 & 3.70 & 5 (2.3)   & 15.66 & 14.84 & 36.39 & 91.04 \\
2018-12-18 & 4.01 & 5 (2.3)   & 16.02 & 15.27 & 32.08 & 75.91 \\
2018-12-23 & 4.16 & 5 (2.3)   & 16.11 & 16.03 & 32.31 & 74.80 \\
2018-12-24 & 4.19 & 5 (2.3)   & 16.12 & 16.06 & 32.59 & 75.16 \\
2018-12-25 & 4.23 & 5 (2.3)   & 16.17 & 15.94 & 31.70 & 72.78 \\
2018-12-26 & 4.26 & 5 (2.3)   & 16.20 & 15.71 & 31.50 & 72.05 \\
2018-12-27 & 4.29 & 5 (2.3)   & 16.21 & 15.33 & 31.85 & 72.56 \\
2018-12-28 & 4.32 & 5 (2.3)   & 16.20 & 15.31 & 32.79 & 74.43 \\
2018-12-29 & 4.35 & 5 (2.3)   & 16.14 & 15.24 & 35.11 & 79.42 \\
2018-12-30 & 4.40 & 5 (2.3)   & 15.59 & 14.96 & 59.47 & 134.04 \\
\enddata            
\end{deluxetable}

\section{Discussion}\label{Discussion}
With imaging observations spanning more than 100 days on either side of perihelion and an average cadence of $3-4$ days between measurements, the observation campaign described in this work provides a unique view into the short and medium term behavior of comet 21P. We will now discuss how these observations compare to observations of comet 21P taken in previous apparitions and what they imply for the physical properties of this comet. 

The steep and asymmetric logarithmic slopes in the dust production of this comet are both in agreement with observations of this comet as during previous apparitions. Significant differences between pre- and post-perihelion behavior of this comet was observed as early as its 1985 apparition \citep{Schleicher1987}, who measured peak \afrho \ values approximately one month before perihelion. Our peak \afrho \ value is measured at a very similar time. During the 1998 apparition, \cite{Lara2003} measured a peak \afrho \ value approximately 2 weeks before perihelion. The times of peak \afrho \ value in both of these papers agree with the measurements presented here, especially given the limited observational data available for the comet during this same time period ($\sim 14-30$ days before perihelion) during all three apparitions. 

 When considering similar ranges of heliocentric distance and fitting to a single power-law model, similar logarithmic slopes and normalizations to those measured in \cite{Blaauw2014} are also measured for this apparition. With the longer baseline of measurements presented here, our newly determined double-exponential model naturally accounts for discrepancies between the measured logarithmic slopes of \cite{Blaauw2014} (which fit pre-perihelion observations) and \cite{Pittichova2008} (which fit post-perihelion observations). Similar behavior was also observed in the comet's water production during its 1998 and 2005 apparitions \citep{Combi2011}. Utilizing observations of the Solar Heliospheric Observatory's (SOHO) \textit{Solar Wind ANisotropies} (SWAN) camera, the water production rate was observed to scale as $\mathrm{r}^{-1.74}_{\mathrm{H}}$ during ingress and $\mathrm{r}^{-11.9}_{\mathrm{H}}$ during egress. In fact, these indices are broadly consistent with the indices measured in this work. This broad overall consistency between water and dust production rates suggests that these two quantities are physically linked to one another.  

Observational data are beginning to support the idea that the dust ejection behavior of comet 21P has not changed significantly over the past few apparitions. Although it is prohibitive to combine the data from every apparition into a single model for comet 21P's dust production rate, it is reassuring from the standpoint of modeling dust ejection from the surface of the comet that a single model is at least qualitatively consistent with all of the apparitions thus observed. We caution, however, that comet 21P has undergone rather dramatic and significant changes in its orbital elements since its discovery. The relative consistency of the past few apparitions does not necessarily imply stability in comet 21P's dust production on longer time scales, especially when considering apparitions earlier than its 1959-1965 orbit, when its non-gravitational forces underwent a sudden change that modified the comet's orbit. 

The results presented here demonstrate the constraining power of a long term monitoring program of a single comet for meteor shower forecasting. Although this observation campaign has measured \afrho \ values for comet 21P at relatively high temporal resolution and over a long period of time on either side of perihelion, poor weather during July, August, and September resulted in relatively few observations in the weeks leading up to perihelion. Unfortunately, these were when \afrho \ values for the comet were at their maximum. Subsequently, this particular time window is the most crucial to accurately model for forecasts of the Draconid meteor shower. We encourage future observation campaigns for comet 21P to place high priority on acquiring images of the comet during this time. Indeed, a modification of the activity slope induced by additional perihelion measurements can modify the time of predicted shower maximum intensity by several minutes, and even hours when the corresponding activity profiles are derived from a small number of simulated meteoroids. 

The ultimate test for these observations will come when Earth is first predicted to encounter meteoroids ejected during this apparition. Simulations conducted over the period 1850-2030 in Egal19 point toward three potential Draconid outbursts over the next decade. In 2025, the Earth might approach a portion of the 2018 trail closer than $5\times10^{-2}$ $\mathrm{AU}$. However, most of the Draconid activity expected in 2025 will rather be produced by meteoroids ejected during the 2005 and 2012 apparitions. From preliminary estimates, an outburst caused by the 2018 trail should almost certainly occur in 2078. We anticipate a follow-up paper comparing the observed Draconid activity of this shower to predictions derived from these observations.

As discussed in Egal19, these measurements of comet 21P's dust production can be immediately coupled to a dynamical stream model in order to produce forecasts for future Draconid shower outbursts. With these measurements in hand, these predictions can only improve.  

\acknowledgments
This work was supported by the NASA Meteoroid Environment Office under contract 80MSFC18C0011 (S.E.), NASA Marshall Space Flight Center's Internship Program (N.M.),  and cooperative agreement 80NSSC18M0046 (A.E.). We thank Dr. William Cooke for his consistent support for this work and providing the resources required to carry out these long-term observations. We thank Herbert Raab and Julio Castellano Roig for making the software programs \textit{Astrometrica} and \textit{FoCAs} publicly available. This work is based on observations taken using the iTelescope network, and we thank the staff of iTelescope for their efforts in developing the telescope network to enable such an observing campaign to be carried out. We finally thank the referee for their careful and throrough review of the initial manuscript. Their comments have certainly improved this paper.

\software{Astrometrica \citep{Astrometrica}, FoCAs \citep{FoCAs}, SExtractor \citep{Bertin1996}, Astropy \citep{Astropy2013,Astropy2018}}

\bibliography{Bibliography}

\begin{thebibliography}{}
\expandafter\ifx\csname natexlab\endcsname\relax\def\natexlab#1{#1}\fi
\providecommand{\url}[1]{\href{#1}{#1}}

\bibitem[{{A'Hearn} {et~al.}(1995){A'Hearn}, {Millis}, {Schleicher}, {Osip}, \&
  {Birch}}]{AHearn1995}
{A'Hearn}, M.~F., {Millis}, R.~C., {Schleicher}, D.~O., {Osip}, D.~J., \&
  {Birch}, P.~V. 1995, \icarus, 118, 223

\bibitem[{{A'hearn} {et~al.}(1984){A'hearn}, {Schleicher}, {Millis}, {Feldman},
  \& {Thompson}}]{AHearn1984}
{A'hearn}, M.~F., {Schleicher}, D.~G., {Millis}, R.~L., {Feldman}, P.~D., \&
  {Thompson}, D.~T. 1984, \aj, 89, 579

\bibitem[{{Astropy Collaboration} {et~al.}(2013){Astropy Collaboration},
  {Robitaille}, {Tollerud}, {Greenfield}, {Droettboom}, {Bray}, {Aldcroft},
  {Davis}, {Ginsburg}, {Price-Whelan}, {Kerzendorf}, {Conley}, {Crighton},
  {Barbary}, {Muna}, {Ferguson}, {Grollier}, {Parikh}, {Nair}, {Unther},
  {Deil}, {Woillez}, {Conseil}, {Kramer}, {Turner}, {Singer}, {Fox}, {Weaver},
  {Zabalza}, {Edwards}, {Azalee Bostroem}, {Burke}, {Casey}, {Crawford},
  {Dencheva}, {Ely}, {Jenness}, {Labrie}, {Lim}, {Pierfederici}, {Pontzen},
  {Ptak}, {Refsdal}, {Servillat}, \& {Streicher}}]{Astropy2013}
{Astropy Collaboration}, {Robitaille}, T.~P., {Tollerud}, E.~J., {et~al.} 2013,
  \aap, 558, A33

\bibitem[{{Astropy Collaboration} {et~al.}(2018){Astropy Collaboration},
  {Price-Whelan}, {Sip{\H o}cz}, {G{\"u}nther}, {Lim}, {Crawford}, {Conseil},
  {Shupe}, {Craig}, {Dencheva}, {Ginsburg}, {VanderPlas}, {Bradley},
  {P{\'e}rez-Su{\'a}rez}, {de Val-Borro}, {Aldcroft}, {Cruz}, {Robitaille},
  {Tollerud}, {Ardelean}, {Babej}, {Bach}, {Bachetti}, {Bakanov}, {Bamford},
  {Barentsen}, {Barmby}, {Baumbach}, {Berry}, {Biscani}, {Boquien}, {Bostroem},
  {Bouma}, {Brammer}, {Bray}, {Breytenbach}, {Buddelmeijer}, {Burke},
  {Calderone}, {Cano Rodr{\'{\i}}guez}, {Cara}, {Cardoso}, {Cheedella},
  {Copin}, {Corrales}, {Crichton}, {D'Avella}, {Deil}, {Depagne}, {Dietrich},
  {Donath}, {Droettboom}, {Earl}, {Erben}, {Fabbro}, {Ferreira}, {Finethy},
  {Fox}, {Garrison}, {Gibbons}, {Goldstein}, {Gommers}, {Greco}, {Greenfield},
  {Groener}, {Grollier}, {Hagen}, {Hirst}, {Homeier}, {Horton}, {Hosseinzadeh},
  {Hu}, {Hunkeler}, {Ivezi{\'c}}, {Jain}, {Jenness}, {Kanarek}, {Kendrew},
  {Kern}, {Kerzendorf}, {Khvalko}, {King}, {Kirkby}, {Kulkarni}, {Kumar},
  {Lee}, {Lenz}, {Littlefair}, {Ma}, {Macleod}, {Mastropietro}, {McCully},
  {Montagnac}, {Morris}, {Mueller}, {Mumford}, {Muna}, {Murphy}, {Nelson},
  {Nguyen}, {Ninan}, {N{\"o}the}, {Ogaz}, {Oh}, {Parejko}, {Parley}, {Pascual},
  {Patil}, {Patil}, {Plunkett}, {Prochaska}, {Rastogi}, {Reddy Janga},
  {Sabater}, {Sakurikar}, {Seifert}, {Sherbert}, {Sherwood-Taylor}, {Shih},
  {Sick}, {Silbiger}, {Singanamalla}, {Singer}, {Sladen}, {Sooley},
  {Sornarajah}, {Streicher}, {Teuben}, {Thomas}, {Tremblay}, {Turner},
  {Terr{\'o}n}, {van Kerkwijk}, {de la Vega}, {Watkins}, {Weaver}, {Whitmore},
  {Woillez}, {Zabalza}, \& {Astropy Contributors}}]{Astropy2018}
{Astropy Collaboration}, {Price-Whelan}, A.~M., {Sip{\H o}cz}, B.~M., {et~al.}
  2018, \aj, 156, 123

\bibitem[{{Bertin} \& {Arnouts}(1996)}]{Bertin1996}
{Bertin}, E., \& {Arnouts}, S. 1996, Astronomy and Astrophysics Supplement
  Series, 117, 393

\bibitem[{{Bessell}(2005)}]{Bessel2005}
{Bessell}, M.~S. 2005, Annual Review of Astronomy and Astrophysics, 43, 293

\bibitem[{{Blaauw} {et~al.}(2014){Blaauw}, {Suggs}, \& {Cooke}}]{Blaauw2014}
{Blaauw}, R.~C., {Suggs}, R.~M., \& {Cooke}, W.~J. 2014, Meteoritics and
  Planetary Science, 49, 45

\bibitem[{Castellano-Roig(2018)}]{FoCAs}
Castellano-Roig, J. 2018, {Fotometria Con Astrometrica}, a Software Tool,
  v3.61, , .
\newblock \url{http://www.astrosurf.com/orodeno/focas/}

\bibitem[{{Combi} {et~al.}(2011){Combi}, {Lee}, {Patel}, {M{\"a}kinen},
  {Bertaux}, \& {Qu{\'e}merais}}]{Combi2011}
{Combi}, M.~R., {Lee}, Y., {Patel}, T.~S., {et~al.} 2011, \aj, 141, 128

\bibitem[{{Copenhagen University} {et~al.}(2006){Copenhagen University},
  {Institute}, {Cambridge}, {Uk}, \& {Real Instituto Y Observatorio de La
  Armada}}]{CMC2006}
{Copenhagen University}, O., {Institute}, A.~O., {Cambridge}, {Uk}, \& {Real
  Instituto Y Observatorio de La Armada}, F. E.~S. 2006, VizieR Online Data
  Catalog, I/304

\bibitem[{{Hosek} {et~al.}(2013){Hosek}, {Blaauw}, {Cooke}, \&
  {Suggs}}]{Hosek2013}
{Hosek}, Matthew~W., J., {Blaauw}, R.~C., {Cooke}, W.~J., \& {Suggs}, R.~M.
  2013, \aj, 145, 122

\bibitem[{{Kronk}(2014)}]{Kronk2014}
{Kronk}, G.~W. 2014, {Meteor Showers} (Springer), doi:10.1007/978-1-4614-7897-3

\bibitem[{Laboratory(2019)}]{JPLHorizons}
Laboratory, N. J.~P. 2019, {{HORIZONS} System}, , .
\newblock \url{https://ssd.jpl.nasa.gov/?horizons}

\bibitem[{{Lara} {et~al.}(2003){Lara}, {Licandro}, {Oscoz}, \&
  {Motta}}]{Lara2003}
{Lara}, L.~M., {Licandro}, J., {Oscoz}, A., \& {Motta}, V. 2003, \aap, 399, 763

\bibitem[{{Mann} \& {von Braun}(2015)}]{Mann2015}
{Mann}, A.~W., \& {von Braun}, K. 2015, Publications of the Astronomical
  Society of the Pacific, 127, 102

\bibitem[{{Marcus}(2007)}]{Marcus2007}
{Marcus}, J.~N. 2007, International Comet Quarterly, 29, 119

\bibitem[{{McFadden} {et~al.}(1987){McFadden}, {A'Hearn}, {Feldman},
  {Bohnhardt}, {Rahe}, {Festou}, {Brandt}, {Maran}, {Niedner}, {Smith}, \&
  {Schleicher}}]{McFadden1987}
{McFadden}, L.~A., {A'Hearn}, M.~F., {Feldman}, P.~D., {et~al.} 1987, \icarus,
  69, 329

\bibitem[{{Monet} {et~al.}(1998){Monet}, {Canzian}, {Harris}, {Reid}, {Rhodes},
  \& {Sell}}]{Monet1998}
{Monet}, D., {Canzian}, B., {Harris}, H., {et~al.} 1998, VizieR Online Data
  Catalog, I/243

\bibitem[{{Pittichov{\'a}} {et~al.}(2008){Pittichov{\'a}}, {Woodward},
  {Kelley}, \& {Reach}}]{Pittichova2008}
{Pittichov{\'a}}, J., {Woodward}, C.~E., {Kelley}, M.~S., \& {Reach}, W.~T.
  2008, \aj, 136, 1127

\bibitem[{Raab(2018)}]{Astrometrica}
Raab, H. 2018, {Astrometrica} Software, Shareware for Research Grade CCD
  Photometry, v4.11.1.442, , .
\newblock \url{http://www.astrometrica.at/}

\bibitem[{{Schleicher} {et~al.}(1987){Schleicher}, {Millis}, \&
  {Birch}}]{Schleicher1987}
{Schleicher}, D.~G., {Millis}, R.~L., \& {Birch}, P.~V. 1987, \aap, 187, 531

\bibitem[{{Schleicher} {et~al.}(1998){Schleicher}, {Millis}, \&
  {Birch}}]{Schleicher1998}
---. 1998, \icarus, 132, 397

\bibitem[{{Sekanina}(1985)}]{Sekanina1985}
{Sekanina}, Z. 1985, \aj, 90, 827

\bibitem[{{Yeomans}(1971)}]{Yeomans1971}
{Yeomans}, D.~K. 1971, The Astronomical Journal, 76, 83

\bibitem[{{Zacharias} {et~al.}(2012){Zacharias}, {Finch}, {Girard}, {Henden},
  {Bartlett}, {Monet}, \& {Zacharias}}]{Zacharias2012}
{Zacharias}, N., {Finch}, C.~T., {Girard}, T.~M., {et~al.} 2012, VizieR Online
  Data Catalog, I/322A

\bibitem[{{Zacharias} {et~al.}(2015){Zacharias}, {Finch}, {Subasavage},
  {Bredthauer}, {Crockett}, {Divittorio}, {Ferguson}, {Harris}, {Harris},
  {Henden}, {Kilian}, {Munn}, {Rafferty}, {Rhodes}, {Schultheiss}, {Tilleman},
  \& {Wieder}}]{Zacharias2015}
{Zacharias}, N., {Finch}, C., {Subasavage}, J., {et~al.} 2015, \aj, 150, 101

\end{thebibliography}

\end{document}